\documentclass{aastex}
\usepackage{emulateapj5}

\usepackage{amssymb}

\slugcomment{Accepted in The Astrophysical Journal: 16 Jan. 2002}
\shorttitle{Resolving the buried starburst in Arp\,299}
\shortauthors{Charmandaris et al.}

\begin{document}


\title{Resolving the buried starburst in Arp\,299}

\author{V. Charmandaris, G.J. Stacey, G. Gull}
\affil{Astronomy Department, Cornell University, Ithaca NY 14853}
\email{vassilis@astro.cornell.edu, stacey@astro.cornell.edu, geg3@cornell.edu}


%\newpage

\begin{abstract}

We present new 37.7\,$\mu$m far-infrared imaging of the infrared
luminous (L$_{\rm IR}\sim 5.16\times10^{11}$L$_\sun$) interacting
galaxy Arp\,299 (= IC\,694 + NGC\,3690). We show that the 38\,$\mu$m
flux, like the 60 and 100\,$\mu$m emission, traces the luminosity of
star forming galaxies, but at considerably higher spatial
resolution. Our data establish that the major star formation activity
of the galaxy originates from a point source in its eastern component,
IC\,694, which is inconspicuous in the optical, becoming visible only
at the near and mid-infrared. We find that IC\,694 is two times more
luminous than NGC\,3690, contributing to more than 46\% of the total
energy output of the system at this wavelength. The spectral energy
distribution of the different components of the system clearly shows
that IC\,694, has 6 times the infrared luminosity of M82 and it is the
primary source responsible for the bolometric luminosity of Arp\,299.
\end{abstract}



\section{Introduction}

One of the major results of the IRAS survey was the discovery of a
class of extremely luminous galaxies (L$\ge$10$^{11}\,$L$_\sun$) which
emit most of their energy in the far-infrared
\citep{Houck84,Soifer89}.  These ultraluminous infrared galaxies
(ULIRGs) have been intensely studied over the past 15 years leading to
significant progress in our understanding of their properties
\citep[see review of][]{Sanders96}. For example, it is now clearly
established that most, if not all, ULIRGs are interacting systems
\citep[i.e.][]{Clements96,Duc97,Borne00}, with ample quantities of
molecular gas and dust surrounding their active central regions
\citep{Solomon97,Gao99}. It appears likely that the mergers
caused the inordinate far-infrared luminosities, either by compressing
the natal ISM and triggering a global starburst, or by triggering
accretion onto a central super massive black hole forming an AGN
\citep{Genzel98,Sanders99,Joseph99}. It is not clear, however,
which mechanism dominates.  This is because massive star formation is
always expected near the central potential well of ULIRGs, but the
spatial resolution at mid- to far-infrared wavelengths, where most of
the energy of those galaxies appears, is rather poor.  For example,
for Arp\,220, the prototypical ULIRG, its two putative nuclei are
separated by $\sim1''$, and most of the molecular gas is well within
the central 10$''$ \citep{Scoville91}. The peak of the far-infrared
spectral energy distribution for Arp 220 though lies near 60\,$\mu$m
where the spatial resolution available from IRAS and ISO is greater
than 30$''$ -- insufficient to resolve the locations of the heating
source(s). Consequently we rely on less direct diagnostics, usually in
the near and mid-infrared, to address the questions of the origins of
the far-infrared luminosity
\citep{Veilleux95,Genzel98,Murphy99,Laurent00,Soifer00}. This approach has 
proven to be extremely powerful and has produced many surprises. One
prime example was the mid-infrared imaging of the Antennae galaxies
(NGC\,4038/39) with the Infrared Space Observatory which revealed that
15\% of its energy at 15\,$\mu$m originated from a star cluster
inconspicuous in the optical and near-infrared \citep{Mirabel98}.
However, while observations in the mid-infrared can much better probe
the dusty cores of IR bright galaxies than visible observations, the
cores can still be extincted in the mid-infrared.  Furthermore, only a
small fraction ($\sim3$\%) of the total luminosity of luminous
infrared galaxies emerges in the mid-infrared \citep{Laurent00}, so
that we are still not directly probing the regions where the
far-infrared flux originates\footnote{For normal late type galaxies it
has been shown that $\sim$15\% of the luminosity is emitted between
5--20$\mu$m \citep{Dale01}}.

Arp\,299\,(=Mrk\,171, VV\,118, IC\,694+NGC\,3690), at a distance of
41\,Mpc, (v$_{\rm hel}$ = 3080\,km\,s$^{-1}$, H$_{\rm
0}$=75\,km\,s$^{-1}$\,Mpc$^{-1}$), is a relatively nearby peculiar
galaxy. The system is dynamically young, its components still
violently interacting leading to the formation of an 180\,kpc (13$'$)
long tidal tail \citep{Hibbard99}. Diffuse emission around the central
regions extends over an area 1.5$'$ in size, while the galaxy itself
consists of IC\,694\,(=Mrk171B = UGC6472 = VV118a) to the east and
NGC\,3690\,(=Mrk171A = UGC6471 = VV118b) nearly 20$''$ to the west
(see Figure~\ref{image}a). Following the nomenclature of
\citet{Gehrz83} and \citet{ww91}, the nucleus of IC\,694 is often
called source A.  NGC\,3690 is resolved to sources B1 and B2 to the
south, with B2 marking the position of its nucleus, and sources C and
C$'$ located $\sim7''$ to the north of B2. It is believed that Arp\,299
results from a prograde-retrograde encounter between a gas rich Sab-Sb
galaxy (IC\,694) and an SBc-Sc galaxy (NGC\,3690) that occured
750\,Myrs ago, and that the system will merge in 20--60\,Myrs
\citep{Hibbard99}. Sources C and C$'$ do not seem to have significant
potential wells to be considered individual galaxies given the lack of
evidence for an underlying concentration of old, red stars
\cite[i.e.][]{Alonso00}. Due to its proximity, Arp\,299 has been
studied extensively ever since it was discovered to have the most
luminous emission lines of any non-Seyfert Markarian galaxy
\citep{Weedman72}. It is among the most X-ray luminous galaxies
\citep[L$_{\rm X}\sim10^{42}$\,ergs\,s$^{-1}$][]{Fabbiano92,Zezas98}
and it is also infrared luminous since, based on its IRAS faint source
catalogue fluxes \citep{Moshir90}, one can calculate\footnote{We use
the standard definition of L$_{\rm
IR}$(8--1000\,$\mu$m)=5.62$\times$10$^5$\,D(Mpc)$^2$\,(13.48{\it
f}$_{12}$+5.16{\it f}$_{25}$+2.58{\it f}$_{60}$+{\it
f}$_{100}$)\,L$_\sun$ \citep[see][]{Sanders96}.}  that L$_{\rm
IR}=5.16\times10^{11}$L$_\sun$. Large quantities of molecular gas with
strong streaming motions are observed in the system, providing ample
fuel for the star formation activity taking place
\citep{Sargent91,Aalto97,Casoli99}. Recent high resolution imaging has
revealed that the morphology of NGC\,3690, which is brighter than
IC\,694 in the optical and near infrared, is quite complex
\citep{Lai99,Alonso00}. Even though there is no compelling evidence
for an active galactic nucleus (AGN) anywhere in the system
\citep[i.e.][]{Augarde85,Armus89,Smith96} it is clear that the
extinction due to dust is high, considerably changing the apparent
morphology of the galaxy as one observes it from the UV and optical to
near- and mid-infrared wavelengths
\citep{Gehrz83,Joy89,Dudley93,Gallais99,Dudley99,Xu00,Soifer01}.

In the present paper we will focus our attention on IC\,694. While
this component appears quite diffuse in the UV and optical, its bulge
becomes obvious as a point source only at near-infrared
\citep{ww91,Lai99}. The description of our observations is presented
in Section 2, the implications of our findings to the current
understanding of Arp\,299 are discussed in Section 3, and finally our
conclusions are summarized in Section 4.

\section{Observations}\label{sect_obs}
\label{sec:obs} 

Continuum images of Arp\,299 at 37.7\,$\mu$m were obtained on 28
February, 1994, using the Kuiper Widefield Infrared Camera
\citep[KWIC, see][]{kwic}.  KWIC is an imaging
spectrometer/spectrophotometer designed for use between 18 and
40\,$\mu$m on the Kuiper Airborne Observatory (KAO). KWIC employs a
128$\times$128 element Si:Sb BIB detector with a 2.73$''$ square pixel
format, so that it oversamples the diffraction limited beam on the KAO
at 38\,$\mu$m ($\lambda/D\sim8''$), yielding resulting
$5.8'\times5.8'$ field of view. KWIC's filters are fully tunable
cryogenic Fabry-Perots, which we fixed at 37.7\,$\mu$m, with a
0.5\,$\mu$m bandwidth for the present observations. The data were
obtained using standard chopping and beam switching techniques, and
calibrated by comparing with images taken of Orion and M\,82 at the
same wavelength.  The system noise equivalent flux density was
$\sim$20\,JyHz$^{-1/2}$, and the beam size was 8.5$''$. The total
integration time was 34 minutes.  For all observations, we dithered
the array to eliminate bad pixels and flat fielded by dividing by an
image of our blackbody calibration source.  Absolute pointing on the
KAO with KWIC was good to $\sim$ 10$''$.

The deep ISOCAM \citep{Cesarsky} broad-band mid-infrared images
presented here \citep[see also][]{Gallais99,Gallais02} were created
using the spectrophotometric observations of Arp\,299 taken on 27
April 1996, after correcting for the transmission of the 7 and
15$\mu$m band-passes (of 2.5 and 5$\mu$m respectively). The estimated
photometric uncertainties of the images are less than 20\%. Details on
this method as well on the reduction and calibration, of the data are
presented by \citet{Laurent00}. The ISOCAM field of view easily
encompasses the whole system with a pixel size of 1.5$''$ and a FWHM
of the point-spread-function (PSF) of $\sim4.5''$ at 15$\mu$m.

\section{Discussion}

\subsection{The 38\,$\mu$m morphology of Arp\,299}
\label{sec:morph} 

\begin{figure} 
\figurenum{1} 
\plotone{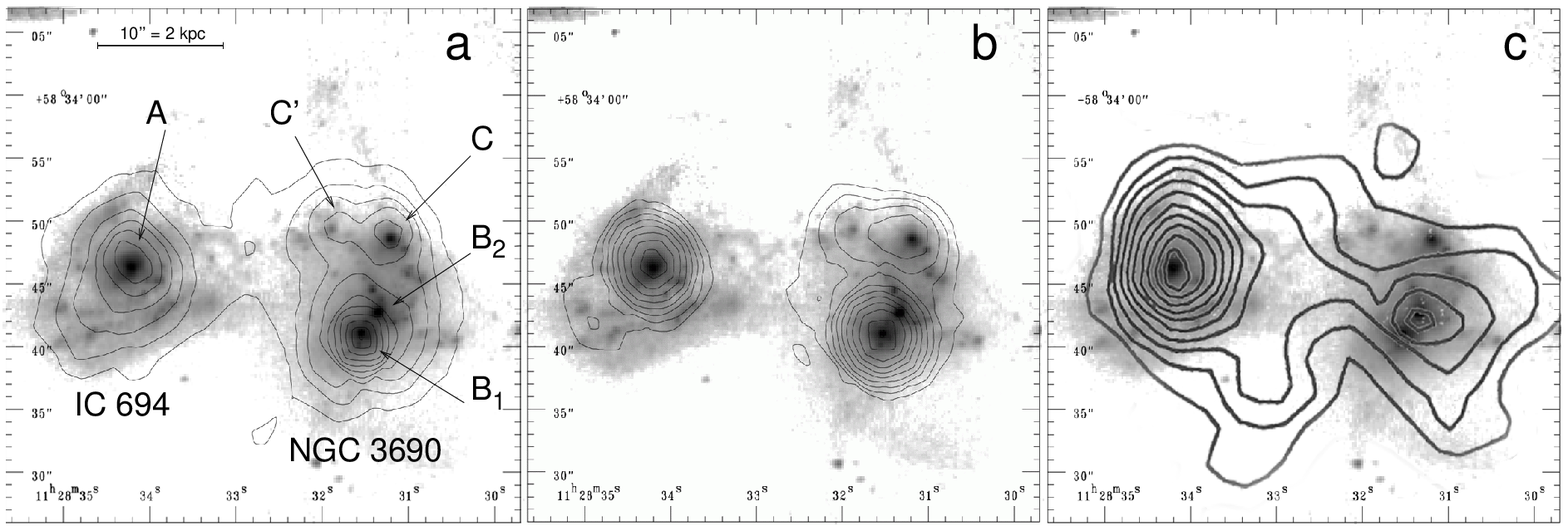}

\caption{ 
a) An ISOCAM 7\,$\mu$m image of Arp\,299 \citep[adapted from
][]{Gallais99} overlayed on an HST/NICMOS 2.2\,$\mu$m image from
\citep{Alonso00}, were the different components of
the galaxy are marked. The 9 contour levels are set with logarithmic
spacing between 1 and 33\,mJy\,arcsec$^{-1}$.  b) Same as in a) but
using the ISOCAM 15\,$\mu$m image as an overlay having set the contour
limits to 6 and 60\,mJy\,arcsec$^{-1}$. c) Our 37\,$\mu$m over the
same HST image. The contour levels are 1.5\,Jy\,beam$^{-1}$ beginning
at 3\,Jy\,beam$^{-1}$ (6\,$\sigma$).\label{image}}
\end{figure}

Our new high spatial resolution two dimensional image of Arp\,299 is
presented in Figure~\ref{image}c. The system is clearly resolved into
its two primary components: source A, the nucleus of IC\,694, and
source B, the nuclear region of NGC\,3690. Source C, the overlap
region to the north of source B, is not apparent as an independent
source, but lies along a ridge of emission extending from source
B. Since the identification of the different components of Arp\,299
has often been challenging and our 8.5$''$ beam is rather large,
extra care was taken in registering the astrometry of our map and it
is therefore useful to describe our method here.  Knowing the pixel
scale and rotation angle of our 38\,$\mu$m image we first compared our
results with the 10\,$\mu$m observations of \citet{Gehrz83} and
\citet{Keto97} which are tied to the highly accurate VLA radio
continuum images. Based on the spectrum of IC\,694, which is discussed
in more detail in the following section, we do not expect the centroid
of source A to change between 10 and 38\,$\mu$m and consequently we
set the PSF fit of our identified source A at the same position as
their measured coordinates for A. Furthermore, we compared our image
to the 7 and 15\,$\mu$m ISOCAM maps of \citet{Gallais99}, revised
versions of which are shown in Figure~\ref{image}a and ~\ref{image}b
and discussed in more detail later. We concluded again that our
positions for the centroid of source A were in agreement, which also
enabled us to tie our 38\,$\mu$m astrometry to the HST/NICMOS J-band
image of \citet{Alonso00} since the latter had been used to bootstrap
the astrometry of the ISOCAM images. Finally, measuring the angular
separation ($23''$) and position angle (-103$\deg$) between our source
A and the second brightest peak in our map, we believe that this
second peak actually coincides with the location of source B1. This
will become more apparent when we discuss the mid-infrared spectral
shape of sources B1 and B2 in the next section. However, given the
fact that the separation between B1 and B2 is just $\sim2''$
\citep[see Fig.4c of ][]{Soifer01}, which is well within our 38\,$\mu$m beam,
we feel that it is more reasonable to consider that our measurement in
that location refers to the flux density of the whole nucleus
B(=B1+B2) of NGC\,3690. Similarly, C and C$'$, which are separated by
$\sim6''$ and are located $\sim9''$ to the north of B1, are not
resolved as point sources. Hence, the value we quote for source C is
the one engulfed in our 38\,$\mu$m beam at that position.

Given the above considerations we find that source A accounts for 46\%
of the total flux density of the galaxy at 38\,$\mu$m. Source B
accounts for 22\% and source C for 8\%. The remaining 24\% is in
``diffuse'' emission over a $45''\times45''$ area covering the
galaxy. Details of our photometry measurements are presented in
Table~\ref{tbl_mir}.

\subsection{The origin of the 38\,$\mu$m emission}
\label{sec:orig} 

Much of the luminosity in infrared bright galaxies emerges at
wavelengths longer than 30\,$\mu$m.  This far-infrared radiation is
``down converted'' optical or UV radiation: dust in the interstellar
medium absorbs optical and UV radiation, heats up, and reradiates the
energy in the far-infrared. For most galaxies, the source of the
optical or UV radiation is starlight, although active galaxies may
have a significant contribution from their active nuclei.  Our
38\,$\mu$m image of the Orion A star formation region
\citep{Stacey95} demonstrated that the 38\,$\mu$m flux arises
from the warm dense photodissociation region at the interface between
the Orion A \ion{H}{2} region and the parent molecular cloud.  For
Orion, the 38\,$\mu$m flux traces the same dust at the 60 and
100\,$\mu$m studies: the observed flux matches the longer wavelength
color temperature and optical depth maps very well.  We expect for
this to hold in general for starburst galaxies as well.  For example,
it is easy to show that for a gray body index of 1.5
(F$_{\nu}=\nu^{\beta}$ B$_{\nu}$) the 38\,$\mu$m flux is proportional
to the far-infrared luminosity to within 50\% for an assumed dust
temperature between 40 and 150\,K.  To first order, then, our
38\,$\mu$m images traces luminosity.

\subsection{The mid- to far-infrared spectrum of Arp\,299}
\label{sec:sed} 

Imaging and spectroscopy at UV, optical, and near-infrared has
revealed that obscuration due to dust throughout Arp\,299 is high. B1,
the nucleus of its eastern component NGC\,3690(=Sources B+C), is
visible from the UV to near-infrared, even though differential
extinction of the various regions of the galaxy (B1, B2, C, and C$'$)
make one or the other source more dominating at a given wavelength
range. It is beyond the scope of this paper to discuss NGC\,3690 in
detail. The latest high resolution analysis of the system by
\citet{Alonso00} using the HST and NICMOS, has demonstrated that it is
IC\,694 the most enshrouded source in the galaxy, in agreement with
earlier work \citep[i.e.][]{Nakagawa89}. Given the high 38\,$\mu$m
flux density of this component in our KAO images, as well as the
findings of \citet{Joy89}, we will focus our attention to its
properties as it appears that IC\,694 is responsible for the largest
fraction of the far-infrared luminosity of Arp\,229.
 
There are several independent clues which lead us to believe that the
dominant starburst of Arp\,299 is buried at the nucleus of
IC\,694. Its UV image (see HST/FOC archive) shows only diffuse
emission with no apparent bright core \citep{Vacca95}. High resolution
HST optical imaging \citep{Malkan98} displays only patchy emission
from several areas of IC\,694, with a dust lane running along the
southeast-northwest direction. Clearly the UV and optical imaging
traces only the surface. The nucleus is not associated with any of the
bright optical features and only becomes visible at near-infrared or
longer wavelengths \citep[see][]{ww91,Lai99,Alonso00}. The radio
continuum and CO line emission \citep{Gehrz83,Aalto97,Soifer01}
coincide with the near infrared peak, confirming that this is the
dynamical center of IC\,694.  It also contains
$3.9\times10^{9}$M$_\sun$ of H$_2$, nearly half of the total observed
in Arp\,299, over an area less than 500\,pc in diameter
\citep{Sargent91} . The high a surface density of molecular gas,
$2.4\times10^{4}\,$M$_\sun$\,pc$^{-2}$ of H$_2$, indicates that
massive star formation can be sustained over $\sim10^{7}$\,yr
\citep{Nakagawa89}.

Source A has a flat radio spectrum and it was first thought that an
AGN was contributing to a considerable fraction of the IR flux
originating from it.  However, as we note from Table~\ref{tbl_mir}, A
does not exhibit an excess of emission in the 3--7\,$\mu$m range
relative to regions B and C. Such a ``hot bump'' would be expected in
an AGN spectrum due to the presence of dust grains associated with the
torus surrounding the active nucleus and heated to near sublimation
temperatures \citep[see the spectrum of NGC\,1068 in][]{LeFloch01}.
Recent work though indicates that the flat radio spectrum of A can be
attributed to free-free absorption and a high supernovae rate of 0.65
yr$^{-1}$ in it.  This rate is almost 5 times that from sources B and
C combined \citep{Alonso00}. The extinction of A is somewhat
uncertain. As it often happens in deeply obscured objects short
wavelength observations merely skim the surface of the star forming
regions and do not probe deep enough to account for the total dust
content. Consequently its quoted value increases as a function of the
wavelength used to measure it: from A$_{V} \sim5$\,mag at
$\sim0.6\mu$m, to $\sim25$\,mag at $\sim2\mu$m, and $\sim25$\,mag
using the CO \citep{Sargent87,Nakagawa89,Alonso00}. The latter value,
however, depends critically on the CO to H$_{2}$ conversion, as well
as the exact CO column density in front of the buried source, both
rather poorly known in this case. Large quantities of dust in A are
also inferred by the deep 9.6$\mu$m silicate absorption feature in
this location -- much more pronounced than in any other component of
the Arp\,299 system \citep{Dudley99}. This result is confirmed by
ISOCAM 5--16\,$\mu$m spectra which show that the 9.6$\mu$m silicate
band of A is indeed nearly saturated with no observable continuum
emission
\citep{Gallais02}.

The overall morphology of the system in the mid-infrared, presented in
Figure~~\ref{image}, is rather intriguing. It is apparent that as we
move from shorter to longer wavelengths, NGC\,3690, which hosts
numerous sites of massive stars and is the major source of the energy
output of the system in the optical and near-infrared, progressively
diminishes in its overall strength presenting a single isolated
unresolved source of emission near the region B1 \citep[see
also][]{Gallais99,Soifer01}. IC\,694 though, displays a steep rise in
its spectrum and at 38\,$\mu$m is more than two times brighter than
NGC\,3690. The ratio of the $f_{38\mu m} / f_{12.5\mu m}$ is 13.8,
$\sim$4 times higher those of regions B and C+C$'$. Unfortunately, we
lack the spatial resolution to examine in detail the behavior of
region C$'$, which according to \citet{Soifer01} displays the highest
$f_{12.5\mu m} / f_{2.2\mu m}$ ratio among the resolved components in
Arp\,299. Consequently, it remains unclear whether this trend
continues in our KWIC data since we can only measure the emission
within one beam at the location of C+C$'$. Even though the difference in
beam sizes and the extent of the old stellar population may conspire
against us we do find that the $f_{38\mu m} / f_{2.2\mu m}$ ratio of
IC\,694 is 890, which is again $\sim$4 times higher than the same
ratio for regions B and C+C$'$. Interestingly at 7 and 15$\mu$m the deep
ISOCAM images reveal that 48\% and 40\% of their respective emission
originates from areas outside regions A, B, and C contrary to just
24\% we find at 38\,$\mu$m. This was suggested by the 3.2\,$\mu$m
images of \citet{Soifer01} and can be attributed to emission from warm
dust and polycyclic aromatic hydrocarbons (PAH) due to star formation
activity in the extended disks and the overlap region between the two
galaxies.

\begin{figure} 
\figurenum{2} 
\plotone{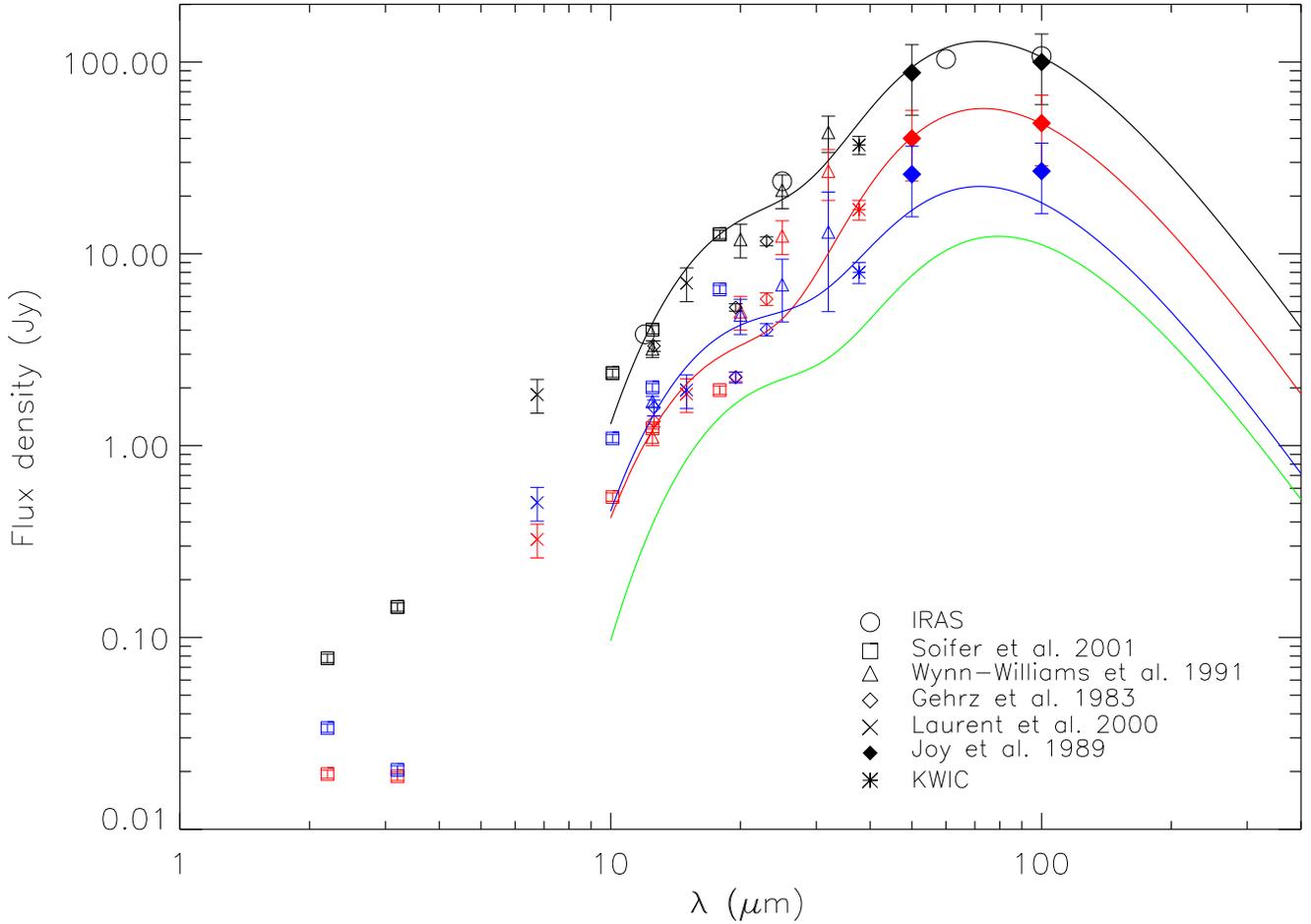}

\caption{ 
The mid- and far-infrared spectral energy distribution of
Arp\,299. The fluxes from region A (IC\,694) are marked with the red
symbols, the ones from region B (NGC\,3690) are shown in blue.  Global
measurements for the whole system are given in black. The gray body
fits (see section \ref{sec:sed}) are also included in the respective
colors. Our fit to the emission from the region C+C$'$ is also displayed
in green. It is evident from the figure that even though NGC\,3690 has
a warmer mid-infrared component, past 20\,$\mu$m the emission from
IC\,694 dominates the observed spectrum of the galaxy. \label{sed}}

\end{figure}

Based on the above information and the data of Table~\ref{tbl_mir} we
plot the spectral energy distribution (SED) for regions A, B, and C of
Arp\,299. We find a good fit to the 10 to 100 $\mu$m fluxes for the
three sources with a two component dust model\footnote{Since multiple
data exist for the same or nearby wavelengths, in our fitting method
we weighted all measurements according to their quoted
uncertainties. For the far-infrared one-dimensional scans of
\cite{Joy89} we conservatively assumed that the flux distribution in
the different components follows our KWIC 38\,$\mu$m image.}, each
with a gray body spectral index of 1.5. The fit is presented in
Figure~\ref{sed} and is valid only for wavelengths longer than
10\,$\mu$m. It is well known that at shorter mid-infrared wavelengths
not only feature emission from PAHs dominates the spectrum, but also
the grains are clearly out of thermal equilibrium. Hence a more
detailed treatment of the underlined physics is necessary
\cite[see][]{Dale01}. The results of our model are presented in
Table~\ref{tbl_fit}. Our model suggests that for component A
T$_{warm}\sim44$\,K, $\tau_{warm}\sim6.8\times10^{-2}$, and T$_{hot}
\sim140$\,K, $\tau_{hot}\sim2.2\times10^{-5}$.  For components B and
C, the model is similar except with smaller optical depth. Based on
our model, we predict that the flux density of A at 100\,$\mu$m is
47\,Jy which suggests\footnote{We use the definition of
M$_{dust}$=4.81$\times$10$^{-12}$ $f_{100}$ D[pc]$^2$ (e$^{143.88/T}$
-1)\,M$_\sun$ from Allen's Astrophysical Quantities.} that its dust
mass is 9.6$\times$10$^6$ M$_\sun$. For a Galactic dust to gas ratio
this would imply $\sim$10$^{9}$\,M$_\sun$ of gas at the center of
IC\,694. Furthermore, if we assume ``outer cloud'' dust, then these
optical depths correspond to a visual extinction, A$_{V} \sim$
250$\times \tau_{38\,\mu m}$, or A$_{V}\sim$17 through the warm dust
at region A.  With a gas column to visual extinction ratio of
2$\times10^{21}$, we also estimate that the gas and dust mass traced
by the 38\,$\mu$m emission at A is $\sim10^{9}$ M$_\sun$. If all our
assumptions are correct then based on the dust properties the
concomitant gas mass is $\sim$7\% of the gas observed
\citep{Hibbard99}. Clearly there is an additional component of cold
dust not traced through its 38\,$\mu$m emission.  However, since the
far-infrared luminosity is nearly entirely in the warm dust component,
and $\sim$75\% of the observed 38\,$\mu$m flux comes from this
component, our analysis confirms that the 38\,$\mu$m flux traces the
far-infrared luminosity.  Integrating the modeled SEDs, we derive
infrared luminosities (F$_{\rm IR}$) of 1.8$\times10^{11}$,
9.4$\times10^{10}$, and 4.4$\times10^{10}$\,L$_\sun$ for A (IC 694),
B(NGC 3690), and C+C$'$(overlap) components respectively. Our model fit
estimates the total infrared luminosity of Arp\,299 to be
4.6$\times10^{11}$\,L$_\sun$, which is 90\% of the amount observed
based only on the IRAS fluxes. One should also note that according to
our model $\sim$30\% of the infrared luminosity of the system is
outside the above defined regions.

\section{Conclusions}

We have obtained high resolution 38\,$\mu$m mid-infrared images of the
peculiar galaxy Arp\,299. Our data clearly shows that despite its
diffuse and quiescent appearance in the UV and optical, IC\,694
harbors a strong point-like source in the mid and far-infrared and is
the dominant source of far-infrared radiation in the system.  Together
with shorter wavelength images, we construct the spectral energy
distribution of the various components of the galaxy.  We find that
IC\,694 is by far the strongest source, with an infrared luminosity of
1.8$\times10^{11}$\,L$_\sun$, or $\sim$40\% of the whole Arp\,299
system.  The infrared luminosity of IC\,694 is 6 times the luminosity
of M\,82, while the one inferred for component C is about 1.5 times
that of M\,82, making it {\em one of the most luminous non-nuclear
starbursts known}.

Our analysis suggests that in order to accurately determine the
starburst activity and infrared luminosity of different regions in
interacting/merging luminous infrared galaxies it is imperative to
obtain good spatial resolution maps covering the 15--40\,$\mu$m range
to better trace the colder dust component. Future instruments such as
IRS and MIPS on SIRTF and FORCAST on SOFIA will provide valuable
information in addressing these problems.

\acknowledgments
We are greatly indebted to the efforts of T. L. Hayward and
H. M. Latvakoski in the design and construction of KWIC and during the
observations. We thank the staff and crew of the Kuiper Airborne
Observatory for their excellent support, especially during the
observatory's challenging final year of operations.  This work was
supported in part by NASA grants NAG2-800 and NAG2-1072.  VC would
also like to thank O. Laurent (Saclay) for help and suggestions on the
optimum analysis of ISOCAM data as well as G. Neugebauer (Caltech) and
E. Le Floc'h (Saclay) for useful discussions.  This research has made
extensive use of the NASA/IPAC Extragalactic Database (NED) which is
operated by the Jet Propulsion Laboratory, California Institute of
Technology, under contract with the National Aeronautics and Space
Administration.

\clearpage

%% Tables should be submitted one per page, so put a \clearpage before

\clearpage

\begin{deluxetable}{lcccccccccccc}
%\rotate
\tabletypesize{\scriptsize} 
\tablecaption{Mid- and Far- Infrared Spectral Energy Distribution of Arp\,299\label{tbl_mir}}
\tablewidth{0pc}
\startdata \\
\tableline 
\tableline \\
Region & \multicolumn{11}{c}{Observed Wavelength}\\ \\
\tableline
 &  2.2\,$\mu$m\tablenotemark{a} & 3.2\,$\mu$m\tablenotemark{a}    
 & 7\,$\mu$m\tablenotemark{b}    & 12.5\,$\mu$m\tablenotemark{a}    
 & 15\,$\mu$m\tablenotemark{b}   & 17.9\,$\mu$m\tablenotemark{a} 
 & 19.5\,$\mu$m\tablenotemark{c} & 20\,$\mu$m\tablenotemark{d}   
 & 23\,$\mu$m\tablenotemark{c}  & 25\,$\mu$m\tablenotemark{d}   
 & 32\,$\mu$m\tablenotemark{d}   & 37\,$\mu$m\tablenotemark{e}\\
  & (mJy) & (mJy) & (mJy) & (mJy) & (mJy) & (mJy) & (Jy) & (Jy)& (Jy)
  & (Jy) & (Jy)& (Jy)\\
\tableline \\

A (=IC\,694) :    &  19.4 & 18.9  & 325   &  1230        &  1860 &
          1950    &  2.29$\pm$0.13 & 5.00  & 5.82$\pm$0.43 & 12.40 &  
          27.00$\pm$8.00 & 17.00 \\
B=B1+B2\tablenotemark{f} :
                 &  22.0+11.7 & 11.7+8.7  & 505   &  2010        &  1951 &
          6520    &  2.27$\pm$0.15 & 4.80  & 4.03$\pm$0.30 & 6.9  & 
          13.00$\pm$4.00 & 8.00\\
C :               &  12.4 &  15.9 & 126   & 436          &   461 &
          3080    &  0.70$\pm$0.12  & 2.10  & 1.81$\pm$0.33 & 2.20  & 
          2.60$\pm$2.30 & 2.95\\
C$'$ :            &  1.12 & 1.99  &  76   & 188           &   232 &
          1100    & \dots & \dots & \dots  & \dots & \dots & \dots\\
Arp\,299\tablenotemark{g}~\,(total) : 
          &  78.1 & 144.2 & 1846 & 4030         &  7037  &
          12650   & 5.26$\pm$0.23 & 11.90  & 11.66$\pm$0.62 & 21.50 & 43.00 & 36.96\\

%% Text for table notes should follow after the \enddata but before
%% the \end{deluxetable}. Make sure there is at least one \tablenotemark
%% in the table for each \tablenotetext.
\tablenotetext{a}{Based on Table~2 and Figure~4e of \citet{Soifer01}. The quoted photometric accuracy is $\sim$5\%.}
\tablenotetext{b}{From \citet{Laurent00} using a 4.5$''$ beam. The photometry is accurate to a 20\% level. }
\tablenotetext{c}{Fluxes densities are for a 5$''$ beam from \citet{Gehrz83}.}
\tablenotetext{d}{From \citet{ww91}. Errors are 20\% unless otherwise specified.}
\tablenotetext{e}{This work.}
\tablenotetext{f}{
We provide the integrated values for source B=B1+B2 at 2.2 and
3.2\,$\mu$m. At longer wavelengths the apertures are centered on B1,
which dominates the emission of B, and any direct contribution due to
B2 is well within the measurement uncertainties.}
\tablenotetext{g}{The integrated IRAS flux densities for the whole galaxy at 12, 25, 60 and 100\,$\mu$m, as presented in the IRAS faint source catalogue, are 3.81, 23.19, 103.7, and 107.4\,Jy respectively \citep{Moshir90}.}

\enddata
\end{deluxetable}

\newpage 

 \clearpage

\begin{deluxetable}{lccccccc}
%\rotate
%\tabletypesize{\scriptsize} 
\tablecaption{Gray body model for Arp\,299\label{tbl_fit}}
\tablewidth{0pc}
\startdata \\
\tableline 
\tableline \\
Region & T$_{warm}$ & $\tau_{warm}$ & T$_{hot}$ & $\tau_{hot}$ 
	&  $\frac{\rm L_{IR {\it warm} }}{\rm L_{IR}}$ 
	&  $\frac{\rm L_{IR {\it hot} }}{\rm L_{IR}}$ & L$_{\rm IR}$ \\
       & (K) & & (K) & & \% & \% & (L$_{\sun}$)\\
\tableline\\
A  & 44 & 6.8$\times$10$^{-2}$ & 140 & 2.2$\times$10$^{-5}$ & 
	88 & 12 & 1.8$\times$10$^{11}$ \\
B=(B1+B2)  & 44 & 2.6$\times$10$^{-2}$ & 135 & 3.5$\times$10$^{-5}$ & 
	67 & 33 & 9.4$\times$10$^{10}$ \\
C+C$'$  & 40 & 2.2$\times$10$^{-2}$ & 120 & 2.8$\times$10$^{-5}$ & 
	67 & 33 & 4.4$\times$10$^{10}$ \\
Arp\,299\,(total) & 44 & 0.15 & 130 & 1.5$\times$10$^{-4}$ & 
	75 & 25 & 4.6$\times$10$^{11}$ \\
\enddata
\end{deluxetable}

%
%  Captions for Figures 
% 
 
\clearpage 

\clearpage 
 \end{document}